\documentclass[%
 aip,jcp,
 amsmath,amssymb,
preprint,%
]{revtex4-1}

\usepackage{graphicx}% Include figure files
\usepackage{dcolumn}% Align table columns on decimal point
\usepackage{bm}% bold math
\usepackage[utf8]{inputenc}
\usepackage[T1]{fontenc}
\usepackage{mathptmx}

\usepackage{hyperref}
\usepackage{graphicx}
\usepackage{float}
\usepackage{grffile}
\usepackage{tabularx} 
\usepackage{array}
\usepackage{framed}
\usepackage{subcaption} 
\newcommand{\eq}[1]{\begin{equation}{#1}\end{equation}}

\begin{document}

\title[Studying polymer diffusiophoresis with Non-Equilibrium Molecular Dynamics]{Studying polymer diffusiophoresis with Non-Equilibrium Molecular Dynamics}
% Force line breaks with \\

\author{S.Ram\'irez-Hinestrosa}
\affiliation{Department of Chemistry, University of Cambridge, Lensfield Road, Cambridge CB2 1EW, United Kingdom}
\author{H. Yoshida}%
\affiliation{LPS, UMR CNRS 8550, École Normale Supérieure, 24 rue Lhomond, 75005 Paris, France}
\affiliation{Toyota Central R\&D Labs., Inc., Bunkyo-ku, Tokyo 112-0004, Japan}

\author{L. Bocquet}
\affiliation{LPS, UMR CNRS 8550, École Normale Supérieure, 24 rue Lhomond, 75005 Paris, France}

\author{D. Frenkel}
\email{df246@cam.ac.uk}
\affiliation{Department of Chemistry, University of Cambridge, Lensfield Road, Cambridge CB2 1EW, United Kingdom}

\date{\today}% It is always \today, today,

\begin{abstract}
We report a numerical study of the diffusiophoresis of short polymers using non-equilibrium molecular dynamics simulations.
More precisely, we consider polymer chains in a fluid containing a solute which has a concentration gradient,
and examine the variation of the induced diffusiophoretic velocity of the polymer chains as the   interaction between the monomer and the solute is varied. 
We find that there is a non-monotonic   relation between the diffusiophoretic mobility and the strength
of the monomer-solute interaction.
In addition we find a weak dependence of the mobility on the length of the polymer chain,
which shows clear difference from the diffusiophoresis of a solid particle. Interestingly, the hydrodynamic flow through the polymer is much less screened than for pressure driven flows. 
\end{abstract}

\maketitle

\section{\label{sec:intro}Introduction}

In a bulk fluid, concentration gradients cannot cause fluid flow. However,  a gradient in the chemical potential of the various components in a fluid mixture, can cause a net hydrodynamic flow in the presence of an interface that interacts differently with the different components of the mixture. Such flow induced by chemical potential gradients is usually referred as diffusio-osmosis (see e.g.~\cite{Anderson1984}). The same mechanism that causes diffusio-osmosis can also drive the motion of a colloid, or other mesoscopic moiety, under the influence of chemical potential gradients in embedding fluid. Clearly, if the mesoscopic particle (say a colloid) is very large compared to the characteristic length scale on which  adsorption or depletion occurs, it is reasonable to use the Derjaguin approximation \cite{Derjaguin1947,Derjaguin1987}, i.e. to describe the colloid-fluid interface as locally flat, and thus estimate the speed of diffusiophoresis. However, the Derjaguin approach is likely to fail if the particles that are subject to phoresis are no longer large compared to the range of adsorption/depletion. There is yet another situation where the Derjaguin approach is obviously questionable, namely in the case of particles that do not have a well-defined surface. One particularly important example is the case of diffusiophoresis of polymers: molecules that have a fluctuating shape  and an intrinsically fuzzy surface. One manifestation of this fuzziness is the fact that the magnitude of the Kirkwood approximation for the  hydrodynamic radius $R_h$ of a long self-avoiding polymer is about 63\% of the radius of gyration $R_g$~\cite{Clisby2016}:
for a smooth sphere, this ratio would be $\approx$ 107\% (the Kirkwood expression for $R_h$ is only an approximation: the point is that the averages are different and that hydrodynamic radius is smaller than for a corresponding solid object). %%%%%%%%
This difference implies that the density inhomogeneity of a self-avoiding polymer results in penetration of hydrodynamic flow fields into its outer ``fuzzy ” layer. In addition, solutes can diffuse through the polymer. This fuzziness clearly makes it difficult to describe a polymer as a solid sphere surrounded with pure liquid, and hence a Derjaguin approach is questionable. The lack of predictive power of the colloidal approximation was pointed out previously by experiments with $\lambda$-DNA by Palacci et al \cite{Palacci2010,Palacci2012}.

There is another factor that makes diffusiophoresis of polymers unusual: since the driving force for diffusiophoresis comes from an excess (or deficit) of solute in the fluid surrounding the polymer, the stronger a solute is attracted to a polymer, the larger this excess will be. However, a strongly binding solute may result in the collapse of the polymer to a compact globule (scaling exponent 1/3). Hence, unlike in the case of colloids, one cannot assume that the size of polymers subject to diffusiophoresis is independent of the polymer-solute interaction. Furthermore, a solute excess/deficit may not be sufficient for the occurrence of diffusiophoresis: if the solutes are strongly adsorbed onto the polymer, they become effectively immobile relative to the polymer, in which case the excess/deficit do not contribute to diffusiophoresis.

In this paper, we report systematic molecular dynamic simulations of diffusiophoretic transport of short polymers. Specifically, we apply non-equilibrium molecular dynamic simulations using a microscopic force acting on each species, and examine the effect of interaction parameters between the monomer and the solute on the induced diffusiophoretic velocity of the polymer. Our simulations indeed reveal a non-monotonic dependence of the phoretic mobility $\Gamma_{ps} $  on  $\epsilon_{ms}$,  the interaction strength between the polymer and solute.
We have investigated the influence of the size of the polymer on its diffusiophoretic mobility. We find a weak polymer-size dependence of the mobility.  We compare these findings with the corresponding theoretical predictions for a colloidal particle.

\section{\label{sec:Thermo-micro}Thermodynamics forces and their microscopic representation}

Conceptually the  most straightforward way of simulating diffusiophoresis would be to carry out a Non-Equilibrium Molecular Dynamics (NEMD) with an imposed concentration gradient, following a procedure similar to Heffelfinger and Van Swol \cite{Heffelfinger1994} and Thompson and Heffelfinger \cite{Thompson1999}. Nonetheless, there are several drawbacks associated with this approach for modeling diffusiophoresis, the most significant being that periodic boundary conditions are incompatible with the existence of constant concentration gradients as advection deforms the concentration profiles (see Supplementary Material). However, in analogy with simulations of systems in homogeneous electrical fields, we can replace the gradient of a chemical potential by an equivalent force per particle that can be kept constant, and therefore compatible with periodic boundary conditions\cite{Yoshida2017,Liu2018}. This field-driven non-equilibrium approach has been often applied in other contexts  \cite{Evans2008}. The idea is to impose a mechanical constraint (i.e an external field) that mimics the effect of the force \cite{Ciccotti1979}.
To see how this approach works in the systems that we study, we first consider diffusio-osmosis in a binary mixture of solvent ($f$) and solute ($s$) particles, which are subjected to a gradient of chemical potential of one of the species (e.g. $s$), and a gradient in the pressure (in bulk fluids in the absence of external body forces, the pressure gradient will typically vanish). We assume that the system is at constant temperature.  Ajdari and Bocquet~\cite{Ajdari2006} derived an expression for the transport matrix $\Gamma$ that relates the fluxes,{\em viz.} the total volume flow $\mathbf{Q}$ and the excess solute flux  $\mathbf{J}_{s}-c_s^B\mathbf{Q}$,  with the gradient of pressure $ -\nabla P$ and the gradient of the chemical potential of one of the species; where $c_s^B$ is the solute concentration in the bulk .  There are only two independent thermodynamic driving forces as, at constant temperature, only two of the three quantities 
$\nabla P$, the solute $\nabla \mu_s$ and the solvent $\nabla \mu_f$ are independent. In fact, it is convenient to define a slightly modified chemical potential gradient  
$\nabla \mu'_s$ by $\nabla \mu'_s$ $\equiv$ $[1+c_s^B/c_f^B]\nabla \mu_s$, in which $c_s^B/c_f^B$ is the ratio between the solute (s) and solvent (f) concentrations in the bulk. With this definition, the linear transport equations can be written as

\eq{
\label{Transport}
\begin{bmatrix}
    \mathbf{Q} \\
    \mathbf{J}_{s}-c_s^B\mathbf{Q}
\end{bmatrix}
=
\begin{bmatrix}
\Gamma_{qq} & \Gamma_{qs} \\
\Gamma_{sq} & \Gamma_{ss} \\
\end{bmatrix}
\begin{bmatrix}
    -\nabla P/T \\
    -\nabla \mu'_s/T
\end{bmatrix},
}

where $\Gamma_{ij}$'s are the Onsager transport coefficients connecting the different fluxes with the thermodynamic driving forces.
In what follows, it is convenient to replace $\nabla \mu_i$ the gradient of the chemical potential on species $i$ by an equivalent external force $\mathbf{F}_{i}^{\mu}$, such that $\mathbf{F}_{i}^{\mu} = -\nabla \mu_i$. This approach to replace the chemical potential gradient by an equal (and opposite) ``color'' force was previously used in the context of diffusion and transport in microporous materials by Maginn {\em et al.}~\cite{Maginn1993,Arya2001}. In the context of diffusio-osmosis,  Liu {\em et al.}~\cite{Liu2018} showed that simulations using color forces yield the same results as those obtained with explicit gradients of concentration. Furthermore, Yoshida {\em et al.}~\cite{Yoshida2017} used the Green-Kubo formalism \cite{Hansen2006} to show that Onsager's reciprocity is also fulfilled. Ganti {\em et al.} have  applied and validated a similar approach in the context thermo-osmosis~\cite{Ganti2017,Ganti2017c}.

Having considered the case of a binary solvent-solute mixture, we now add a third component, namely the polymer, to the system. Again, not all chemical potential gradients are independent, as it follows from the Gibbs-Duhem equation:
\eq{VdP=SdT+\sum_i N_i d\mu_i.} 
In what follows, we assume there are no global pressure and thermal gradients in the system. As a consequence, we can write:
\eq{\mathbf{F}_{p}^{\mu}=-(\mathbf{F}_{s}^{\mu}N_s+\mathbf{F}_{f}^{\mu}N_f), \label{eq:monomer_force}}
where $\mathbf{F}_{p}^{\mu}$,$\mathbf{F}_{s}^{\mu}$,$\mathbf{F}_{f}^{\mu}$ denote the equivalent forces on the polymer, solute and solvents mimicking the corresponding chemical potential gradients. $N_s$, $N_f$ refer to the total number of solutes and solvents in the system as a whole. This equation simply expresses the fact that there can be no net external force on  the fluid: if there were, the system would accelerate without bound, as there are no walls or other momentum sinks in the system. 
In simulations, it is convenient to work with a force per monomer, rather than a force on the center-of-mass of the polymer: $\mathbf{F}_{m}=\mathbf{F}_{p}^{\mu}/N_m$, where $N_m$ denotes the number of beads in the polymer.
Equation~\eqref{eq:monomer_force} establishes a connection between all  chemical potential gradients (or the corresponding microscopic forces), which must be balanced throughout the system as phoretic flow cannot cause bulk flow.

We will now compute the rate of polymer diffusiophoresis using Eq.~\eqref{eq:monomer_force} as our starting point.

\section{Molecular Dynamics (MD) simulations }

We performed non-equilibrium Molecular Dynamics (NEMD) simulations using LAMMPS \cite{Plimpton1995}. In most simulations,  particles interact via a 12-6 Lennard-Jones potential (LJ) $V_{LJ}(r)=4\epsilon^{LJ}_{ij}[(\sigma^{LJ}_{ij}/r)^{12}-(\sigma^{LJ}_{ij}/r)^6]$ shifted and truncated at $r=r_c$, such that 
\begin{equation}
    V_{TS}(r)=
    \begin{cases}
      V_{LJ}(r)-V_{LJ}(r_c), & \text{if}\ r \le r_c \\
      0, & \text{otherwise}.
    \end{cases}
  \end{equation}

The indices $i$ and $j$ denote the particle types in our simulations: solutes ($s$), solvents ($f$) and monomers ($m$). To keep the model as simple as possible, 
we assume that in the bulk the solute and solvent behave as an ideal mixture. We therefore choose the same Lennard-Jones interaction for the particle pairs $ss$, $sf$, $ff$ with $\epsilon^{LJ}_{ij}=\epsilon_0$ and $\sigma^{LJ}_{ij}=\sigma_0$. We use these same parameters also  for the monomer-solvent interaction $ms$. However, the monomer-solute interaction strength $\epsilon^{LJ}_{ms}$ was
varied to control the degree of solute adsorption or depletion around the polymer. Yet, we kept $\sigma^{LJ}_{ms}$ equal to $\sigma_0$.
For the monomer-monomer interaction, we use a  purely repulsive Weeks-Chandler-Andersen potential~\cite{Weeks1971}, i.e. a Lennard-Jones potential truncated and shifted  at the minimum of the LJ potential, $r_c = 2^{1/6} \sigma_0$.
 For all other interactions, $r_c = 2.5 \sigma_0$. Finally, neighboring monomers are connected by a finite extensible, nonlinear elastic (FENE) anharmonic potential $U_{FENE}(r)$, \cite{Bishop1979,Dunweg1993} 
\eq{\label{eq:FENE}U_{FENE}(r)=-\frac{kR_0^2}{2}\ln\left[1-\left(\frac{r}{R_0}\right)^2\right],}
with $k=7 \epsilon_0/\sigma_0^2$ and $R_0=2 \sigma_0$. In what follows, we use the mass $m_0$ of all the particles ($s$,$f$ and $m$) as our unit of mass and 
we set our unit of energy equal to $\epsilon_0$, whilst our unit of length is equal to $\sigma_0$, all other units are subsequently expressed in term of these basic units. As a result, forces are expressed in units
$\epsilon_0/\sigma_0$, and our unit of time is  $\tau\equiv\sigma_0 \sqrt{m_0/\epsilon_0}$.

\subsection{Equilibration}

\begin{figure}[H]
\centering
\includegraphics[scale=0.4]{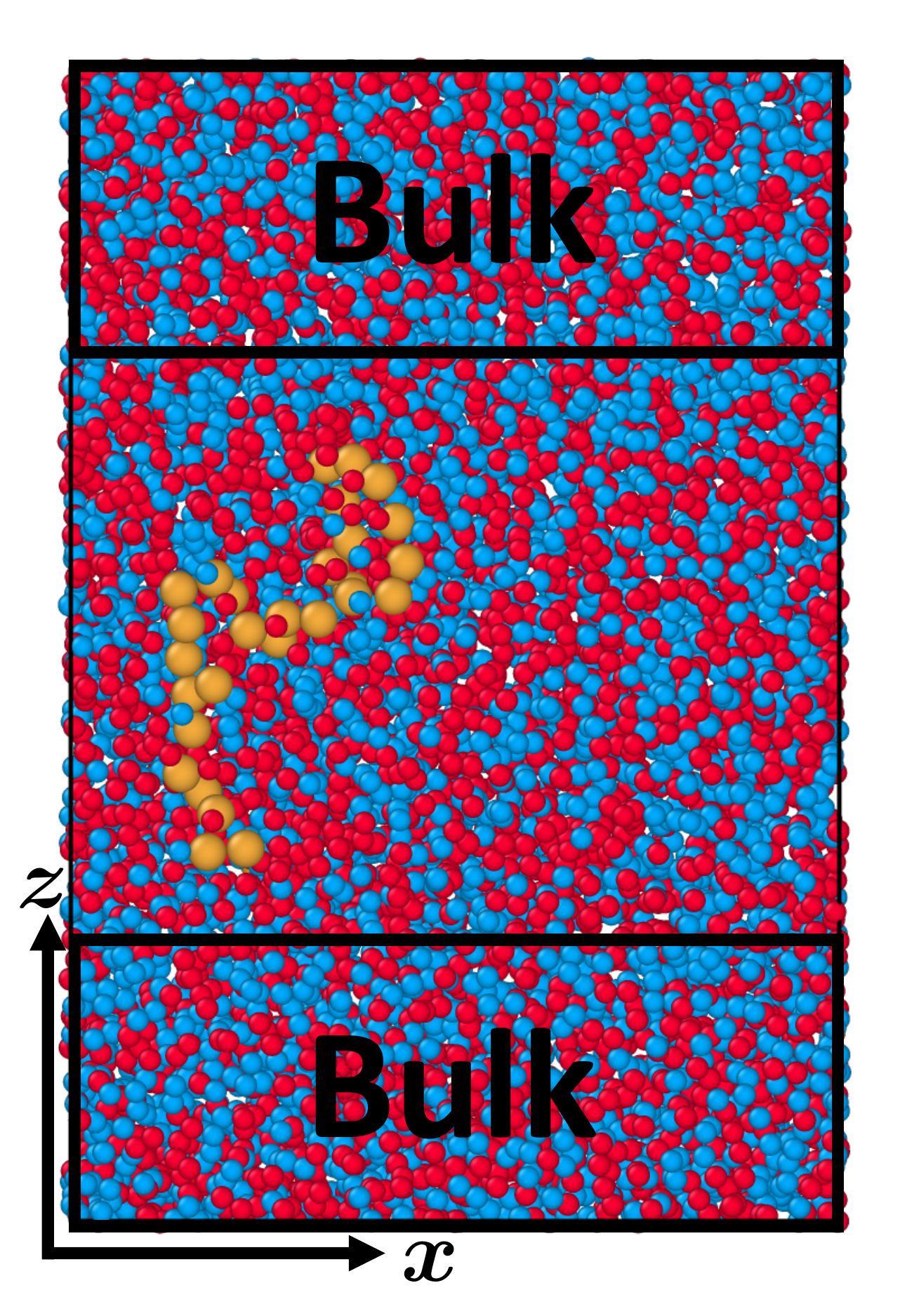}
\caption{Simulation box showing solutes (red), solvents (blue) and monomers (orange). The bulk regions are shown inside black boxes. In the bulk, the solute and solvent concentrations are assumed to be unperturbed by the presence of the polymer.}  
\label{fig:bulk}
\end{figure}

We studied the diffusiophoresis of a single polymer chain composed of 30 monomers, $N_m=30$, suspended in an equimolar ideal mixture of solute and solvent molecules. 
The initial simulation box dimensions were $L_x=20 \sigma_0$, $L_y=20 \sigma_0$, $L_z=30 \sigma_0$ and the number of fluid particles was 8748. 
After equilibration, chemical potential gradients were applied along the $x$-axis. We distinguished two types of domains in the simulation box: one periodically repeated domain with width $20\sigma_0$ in the $z$-direction centered around the polymer's center of mass in the $z$-coordinate. The remainder of the system ( a domain with width $10\sigma_0$), contains only bulk fluid (see Fig \ref{fig:bulk} -- note that because this figure is centered around the polymer, one half of the bulk domain is shown above and one half below the polymer domain). We verified that the composition of the mixture in this bulk domain is not influenced by the presence of the polymer in the other domain (See Supplementary material). The system is periodically repeated on each direction.\\

Our aim was to carry out simulations under conditions where the composition of the bulk fluid was kept fixed, even as we varied the monomer-solute interaction. Moreover, we prepared all systems at the same hydrostatic pressure. Therefore, we performed NPT simulations using a Nos\'{e}-Hoover thermostat/barostat \cite{Hoover1985}. The equations of motion were integrated using a velocity-Verlet algorithm with time step  $\ t= 0.005\tau$. After the relaxation of the initial configuration, the box was allowed to fluctuate in the $y$ direction, fixing $k_BT/\epsilon_0=1.0$ and $P\sigma_0^3/\epsilon_0=1.0$ for $2 \times 10^4$ steps.

During the NPT equilibration, fixing the bulk concentration of the liquid requires a careful protocol, in particular in cases where the solute binds strongly to the polymer.  In our simulations, we accelerated the equilibration of the solute adsorption on the polymer by attempting to swap solvent and solute molecules $10^4$ times for every MD step throughout the simulation box. Simultaneously, we swapped solutes and solvents in the bulk, to ensure that adsorption on the polymer does not deplete the solute concentration in the bulk. To this end, we swapped solutes and solvents in the bulk  every 200 steps  such that the bulk solute concentration remained fixed at $c^B_s\approx 0.376$. Note that these swaps were only carried out during equilibration.

\subsection{Field-driven simulation}
Having prepared a system of polymer in a fluid mixture with a pre-determined pressure and bulk composition, we now consider the effect of chemical potential gradients on the phoretic motion of the polymer in NVT simulations. As discussed above, we represent the chemical potential gradients by equivalent external forces that are compatible with the periodic boundary conditions. Importantly, the forces are chosen such that a) there is no net force on the system as a whole and b) there is no net force on the bulk solution away from the polymer.  These two conditions imply that there is only one independent force that can be defined in the system. In the present case, we chose to fix the force on the solutes ${F}_{s}^{\mu}$, which was varied between   $0$ and $0.1$ $\epsilon_0 /\sigma_0$ for different runs. During all the field-driven simulations, we employed a dynamical definition of the bulk and polymer domains such that the $z$-coordinate of the center of mass of the polymer is always in the middle of the polymer domain.
This procedure ensures that the ``bulk'' region remains unperturbed by the polymer. Having specified the force on the solute, the force on the solvent particles follows from mechanical  equilibrium in the bulk in Eq.~\eqref{eq:monomer_force} (see Fig. \ref{fig:bulk}):
\eq{\mathbf{F}_{s}^{\mu} N_s^B +\mathbf{F}_{f}^{\mu}  N_f^B  = 0, \label{eq:bulk_equilibrium}}
where $N_s^B$, $N_f^B$ denote the number of solutes and solvent in the bulk region. Once the forces in the bulk have been specified, the phoretic force on the polymer $\mathbf{F}_{p}^{\mu}$ is obtained by imposing force balance on the system as a whole (Eq.~\eqref{eq:monomer_force}).

Due to the finite size of the bulk domain there are inevitably fluctuations in the composition of this domain. These fluctuations would lead to unphysical velocity fluctuations in the bulk (unphysical because in the thermodynamic limit this effect goes away). These velocity fluctuations would contribute to the noise in the observed phoretic flow velocity. To suppress this effect, we could either adjust the composition in the bulk domain at every time step, or adjust the forces  on solute and solvent ($\mathbf{F}_{p}^{\mu}$ and $\mathbf{F}_{f}^{\mu}$) such that the external force on the bulk domain is always rigorously equal to zero. We opted for the latter approach, because particle swaps would affect the stability of the MD simulations.

\section{Results and discussion}

\subsection{Phoretic velocity}

\begin{figure}[H]
\centering
\includegraphics[scale=0.45]{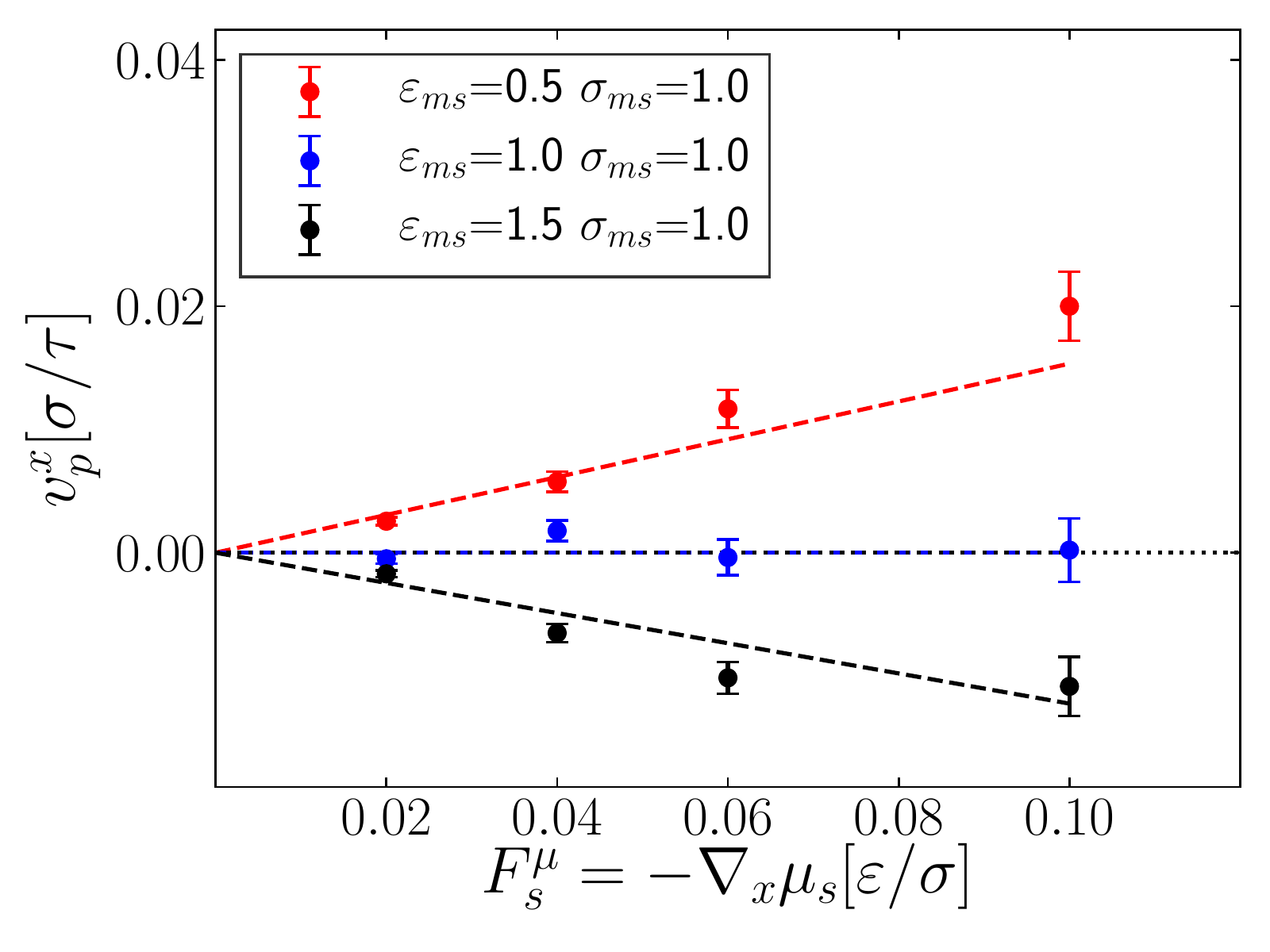}
\caption{Polymer velocities in the direction of the gradient for different LJ interactions ($\epsilon^{LJ}_{ms},\sigma^{LJ}_{ms}$) vs the force applied on the solute particles.}  
\label{fig:velocities}
\end{figure}

In Fig \ref{fig:velocities} the polymer velocities in the direction of the gradient $v^x_p$ are plotted for three different pair of LJ parameters. When there is adsorption of solutes around the polymer ($\epsilon^{LJ}_{ms}=1.5)$, the polymer follows the gradient, migrating towards regions where the solute concentration is higher.  Conversely, when there is depletion ($\epsilon^{LJ}_{ms}=0.5)$ the polymer will move in the opposite direction.  As a null check, we also performed simulations for the case where the $\epsilon^{LJ}_{ms}=\epsilon^{LJ}_{mf}$. In that case, there should be no phoresis, as is indeed found in  the data shown in  Fig.~\ref{fig:velocities}. The inversion of the velocity depending on the sign of the monomer-solute interaction is expected on the basis of irreversible thermodynamics~\cite{Anderson1984} and has previously been observed in simulations of for nano-dimers, using hybrid molecular dynamics-multiparticle collision (MD-MPC) dynamics \cite{Ruckner2007,Tao2008}. 

The figure also shows that our simulations appear to be in the linear regime, as  the magnitude of the phoretic velocity increases linearly with the strength of the applied field.

%%%%%%%%%%%%%%%

\subsection{Mobility dependence on the interaction}
\begin{figure}[H]
\centering
\includegraphics[scale=0.45]{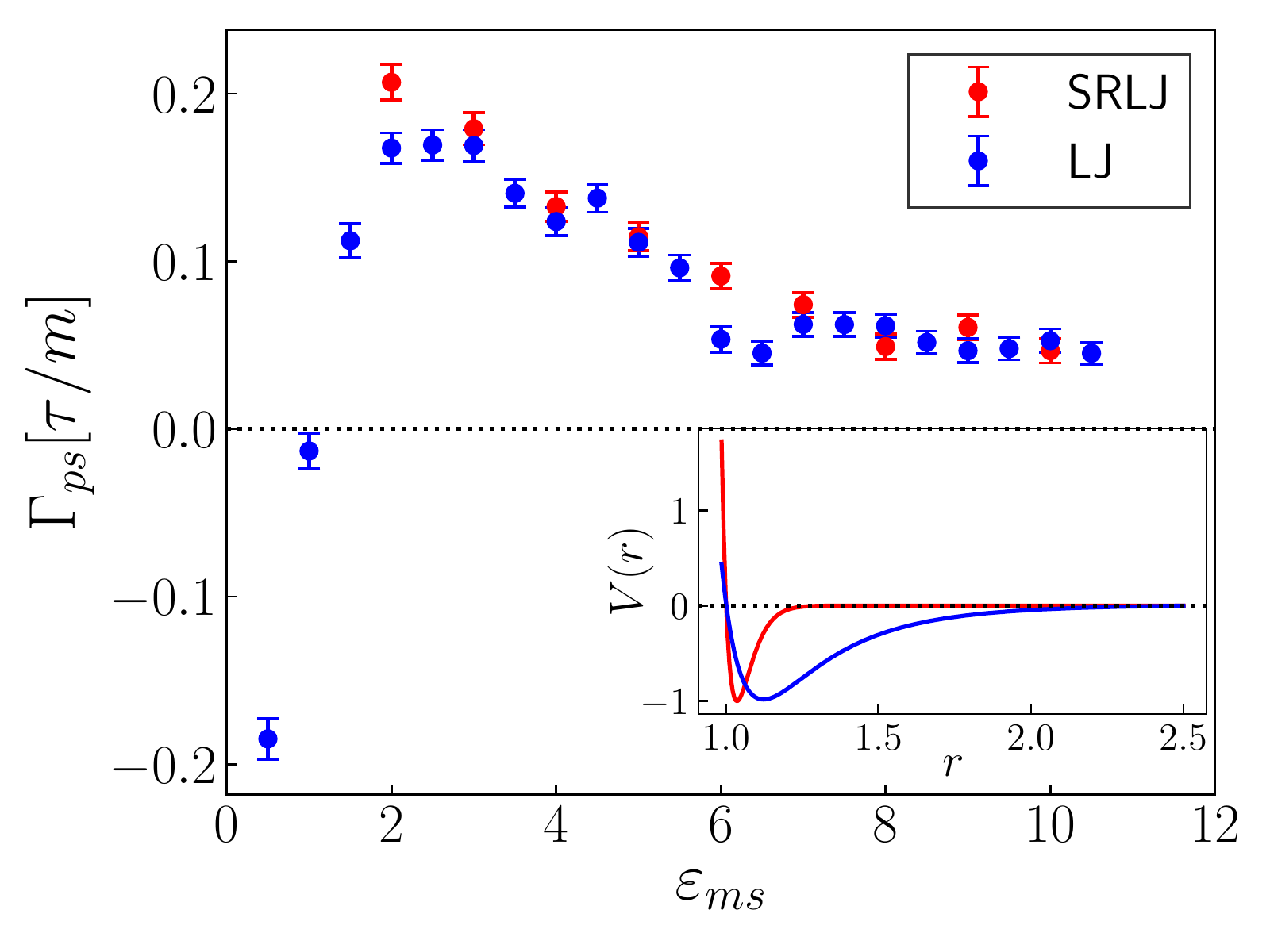}
\caption{Mobilities for different LJ interaction energies ($\epsilon_{ms}$). All the simulations were performed keeping the thermodynamic conditions in the bulk constant ($T$,$P$, $c_s^B$). The insert shows a LJ potential for $r_c=2.5$, $\epsilon^{LJ}=1$ and $\sigma^{LJ}=1$ and a SRLJ potential, showing the narrow range of the monomer-solvent interaction.}  
\label{fig:Mobility_frenkel}
\end{figure}

The mobility $\Gamma_{ps}$ of a polymer moving under the influence of a gradient in the solute chemical potential  is defined through:
\eq{\mathbf{{v}}_p^x=\Gamma_{ps}\mathbf{\nabla}_x \mu_s.}
We can compute $\Gamma_{ps}$ as a function of the polymer-solute interaction strength from the slope of the ${v}_p^x$ {\em vs.}\ $\nabla_x \mu_s$ plots, such as the ones shown in  Fig \ref{fig:velocities}. This procedure allows us to obtain $\Gamma_{ps}$ as a function of the monomer-solute interaction strength
 $\epsilon^{LJ}_{ms}$. We stress that, whilst we determine $\Gamma_{ps}$ by varying $\nabla_x \mu_s$, we keep the bulk composition of the mixture fixed (as well as the temperature and the pressure). The resulting  relation between  $\epsilon^{LJ}_{ms}$ and  $\Gamma_{ps}$ is shown in Fig \ref{fig:Mobility_frenkel}.
 As expected,  $\Gamma_{ps}$ is  linear in $\epsilon^{LJ}_{ms}$ when $\epsilon^{LJ}_{ms}/\epsilon^{LJ}_{mf}$ is close to one. However, as the monomer-solute interaction gets stronger, $\Gamma_{ps}$ saturates, and subsequently decays with increasing  $\epsilon^{LJ}_{ms}$.

The observed decrease of $\Gamma_{ps}$ for large values of  $\epsilon^{LJ}_{ms}$ suggests that when solute particles bind strongly to the polymer, they become effectively immobilised and hence cannot contribute to the diffusio-osmotic flow through and around the polymer. This argument would suggest  that the diffusiophoretic velocity should vanish as $\epsilon^{LJ}_{ms}$ becomes much larger than the thermal energy. However, that does not seem to be the case: rather $\Gamma_{ps}$ seems to level off at a small but finite value. This suggests that not all fluid particles involved in the phoretic transport are tightly bound to the polymer. 
One obvious explanation could be that the LJ potential that we use is sufficiently long-ranged to interact with solute particles that are in the second-neighbour shell around the monomeric units of the polymer. To test whether this is the case, we repeated the simulations with a shorter-ranged short-ranged Lennard-Jones-like potential (SRLJ)\cite{Wang2019}  that has a smaller cut-off distance ($r_c=1.6$ ) where the potential and its first derivative vanish continuously. In the insert of Fig. \ref{fig:Mobility_frenkel}  this  narrower potential is shown compared with the standard truncated and shifted  LJ potential with $r_c=2.5$, $\epsilon^{LJ}=1$ and $\sigma^{LJ}=1$.
In our  simulations we only used the SRLJ potential for the monomer-solute interactions. For all other interactions we still use the standard 
 LJ potential.
 
\begin{figure}[H]
\centering
\begin{subfigure}{0.5\linewidth}
  \centering
\caption{}
  \includegraphics[width=1\linewidth]{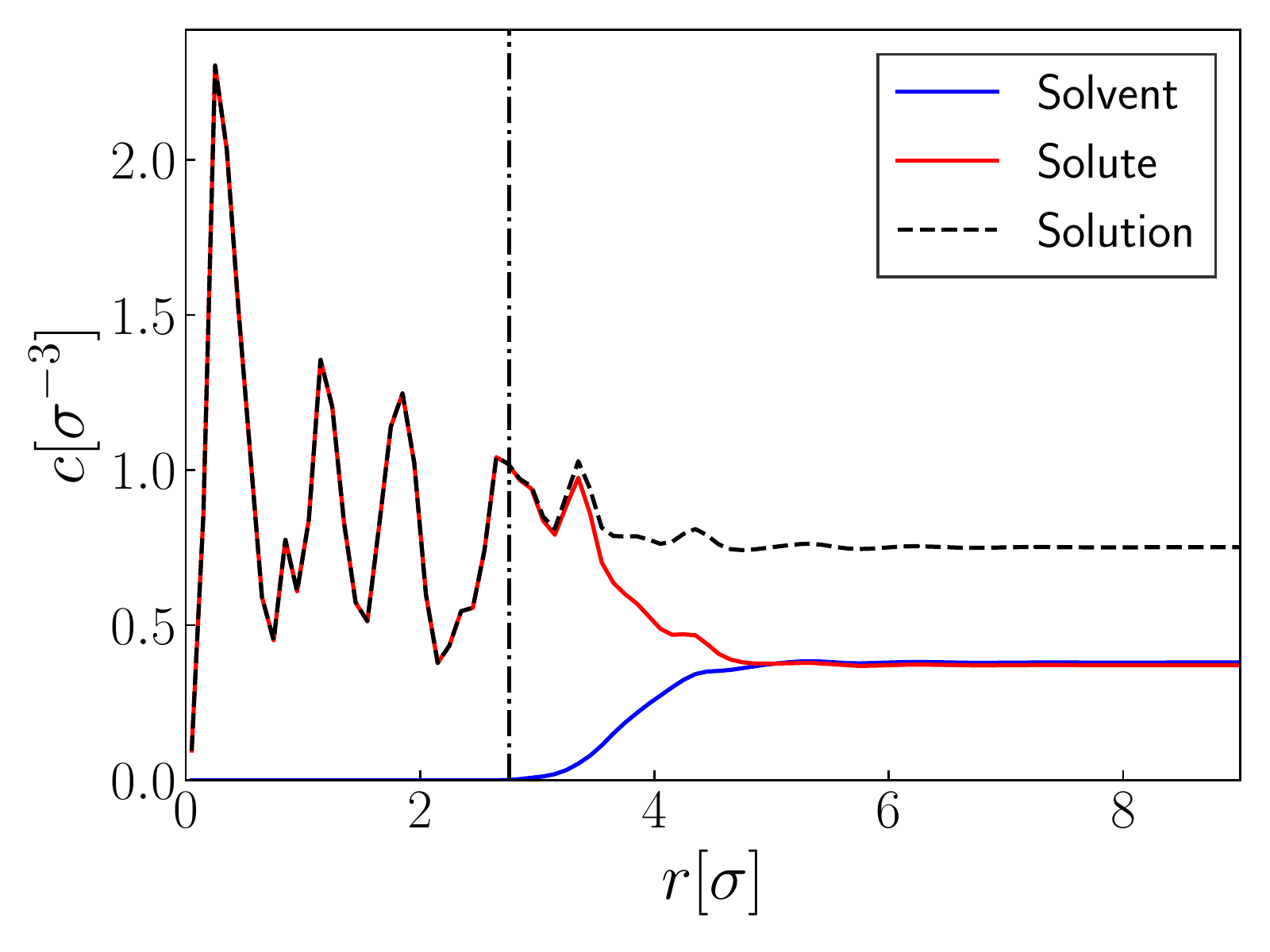}
  
  \label{fig:structuringa}
\end{subfigure}%
\begin{subfigure}{0.5\textwidth}
  \centering
  \caption{}
  \includegraphics[width=1\linewidth]{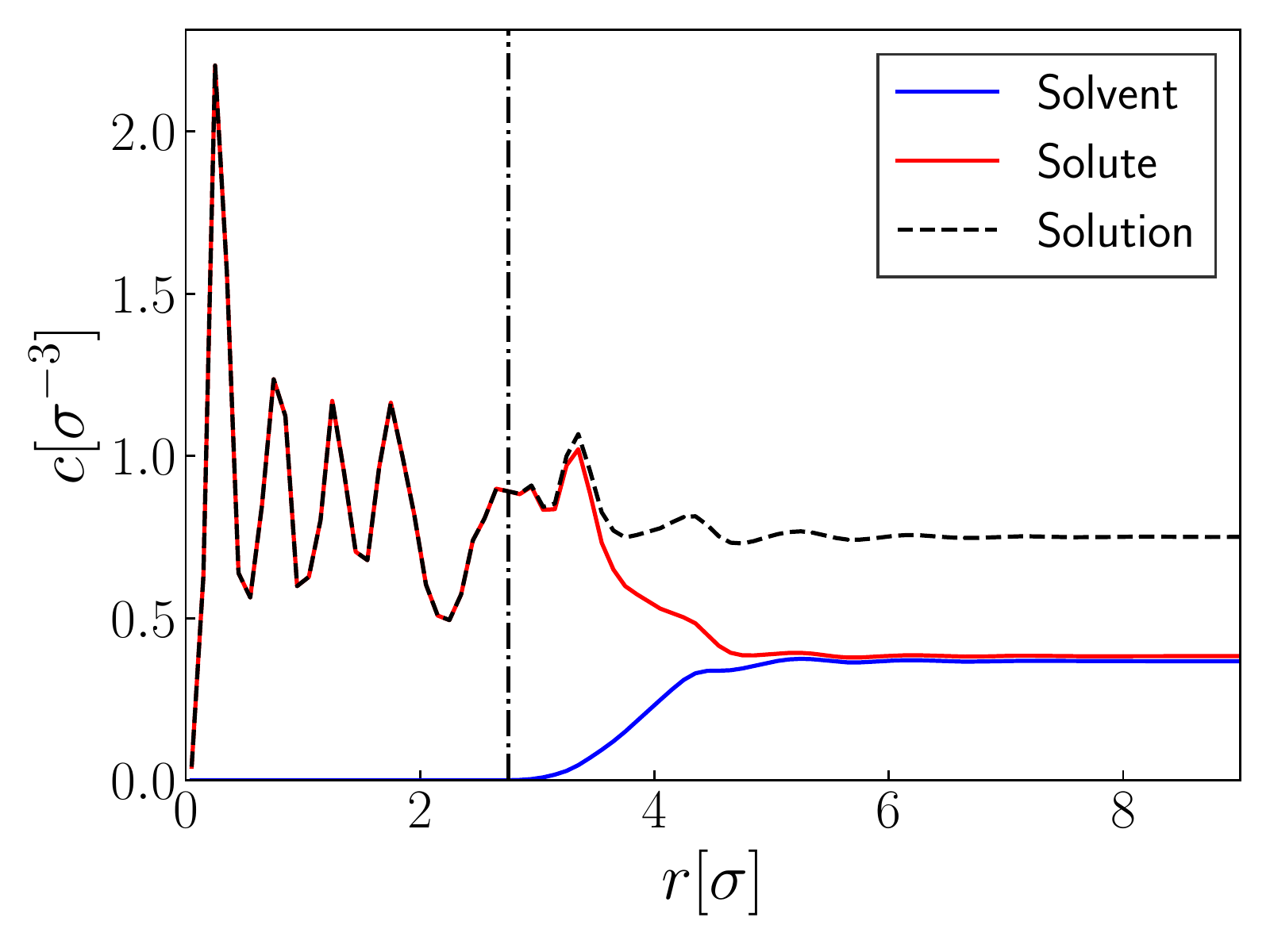}

  \label{fig:structuringb}
\end{subfigure}
\caption{Distribution of solutes, solvents and the total solution measured from the center of mass of the polymer for (a) LJ, (b) SRLJ. The vertical line represents the hydrodynamic radius $R_h$ of the polymer, in both cases $\epsilon^{LJ}_{ms}=8.0$. Mobile particles in the heterogeneous region outside the collapsed polymer coil contribute to the diffusio-osmotic flow in a similar way for both ranges of interaction. }
\label{fig:structuring}
\end{figure}

Figure \ref{fig:Mobility_frenkel} shows a comparison of the results obtained with the LJ and the SRLJ potentials. Interestingly, even with the short-ranged monomer-solute interaction for which next-nearest neighbour interactions are excluded, $\Gamma_{ps}$ still does not decay to zero at large $\epsilon^{LJ}_{ms}$. This suggests that the phoretic force is not just probing the excess of solute particles that are directly interacting with the polymer, but also the density modulation of solutes (and solvent) that is due to the longer-ranged structuring of the mixture around the polymer coil. In Fig. \ref{fig:structuring}, we show an extreme case ($\epsilon^{LJ}_{ms}=8.0$) where the polymer has collapsed and particles within a hydrodynamic radius $R_h$ from the center of mass do not contribute to phoresis as they are tightly bound. In contrast, particles in the structured liquid layer further away from the center of the polymer ($r>R_H$) are mobile and can therefore contribute to the diffusio-osmotic flow.

\subsection{Scaling of the phoretic mobility with the length of the polymer}
For colloidal particles with a radius much larger than the range of the colloid solute interaction, the diffusiophoretic mobility is independent of the colloidal radius~\cite{Anderson1982}. As the diffusion of a polymer in a fluid is often described as that of a colloid with an equivalent ``hydrodynamic radius'' $R_h$, one might be inclined to assume that the diffusiophoretic mobility of a sufficiently large polymer might also be size independent. To our knowledge, this size dependence has not been tested in simulations. However, experiments by  Rauch and K\"{o}hler~\cite{Rauch2005} showed that thermophoretic mobility of polymers varies with the molecular weight $M_w$ for short polymers (fewer than 10 monomers), but  very little for longer polymers (10-100 monomers).\\

For colloids, Anderson~\cite{Anderson1982} derived an expression for the diffusiophoretic mobility of colloids in the case where the interfacial layer thickness $L$ is smaller, but not much smaller than the radius $a$ of the colloid. Introducing the small parameter $\lambda\equiv L/a$, Anderson derived the following asymptotic expression for the diffusiophoretic velocity $v$ of a colloidal particle:

\eq{\label{eq:dpvelocity}v=v_0\left[ 1-  \frac{(K+H)}{L}\lambda+ \mathcal{O}(\lambda^2) \right].}

In this approximation, the first term corresponds to the Derjaguin limit $L\ll a$:

\eq{\label{eq:flatplate}v_0=\frac{\alpha}{\beta \eta}L^*K,}
where $\alpha$ is the magnitude of the concentration gradient, $\beta=1/(k_BT)$, $\eta$ is the shear viscosity and $\phi$ is the  potential of mean force experienced by  solutes at a distance $y=r-a$ from the surface of the colloid. $K$, $L^*$, $H$ are proportional to moments of the excess solute distribution $c_s(y)=c_s^B \exp[{-\beta \phi(y )}]$,
\eq{\label{eq:zeroth}K=\int_0^\infty[\exp[{-\beta \phi(y ) } ]-1] dy,}
\eq{\label{eq:first}L^*=\frac{\int_0^\infty y[\exp[{-\beta \phi(y )}]-1]dy}{K},}
\eq{\label{eq:second}H=\frac{\displaystyle \int_0^\infty \frac{1}{2}y^2[\exp[{-\beta \phi(y) } ]-1] dy } { \displaystyle \int_0^\infty y[\exp[{-\beta \phi(y) } ]-1] dy }.}
The corrections terms in Eq. \eqref{eq:dpvelocity}  account for the effect of the curvature of the particle. All the above equations apply to the case where there is no hydrodynamic slip on the surface of the colloid. However, if solute particles are strongly adsorbed to the colloid, they become immobile and the result is simply that the surface of no slip, and hence the effective colloidal radius, increases. Eq.~\eqref{eq:dpvelocity} was derived assuming no-slip boundary conditions to solve the Navier-Stokes equation. However, Ajdari and Bocquet \cite{Ajdari2006} showed that a correction due to the hydrodynamic slip captures the transport enhancement at interfaces. Including the amplification factor due the surface slip, for moderate adsorption or depletion of solutes, the corrected diffusiophoretic velocity $v'$ reduces to:  

\eq{\label{eq:dpvelocity_corrected}v'=v\left(1+\frac{b}{L}\right).}
where $b$ is the hydrodynamic slip length.

To study the dependence of the phoretic motion of a polymer on the number of monomers of the chain $N_m$, simulations were performed for a range of $N_m$ from $5$ to $60$. As our polymers are fully flexible (but self-avoiding) a chain of 60 beads corresponds to a medium-sized polymer. The simulation box was scaled accordingly with the Flory exponent for a polymer in a good solvent $\nu\approx 0.6$
thus ensuring that the chain could not overlap with its periodic images.  All the NEMD simulations were carried out for $10^8$ time steps. 

\begin{figure}[H]
%this plot was performed with plotall.py
\centering
\includegraphics[scale=0.6]{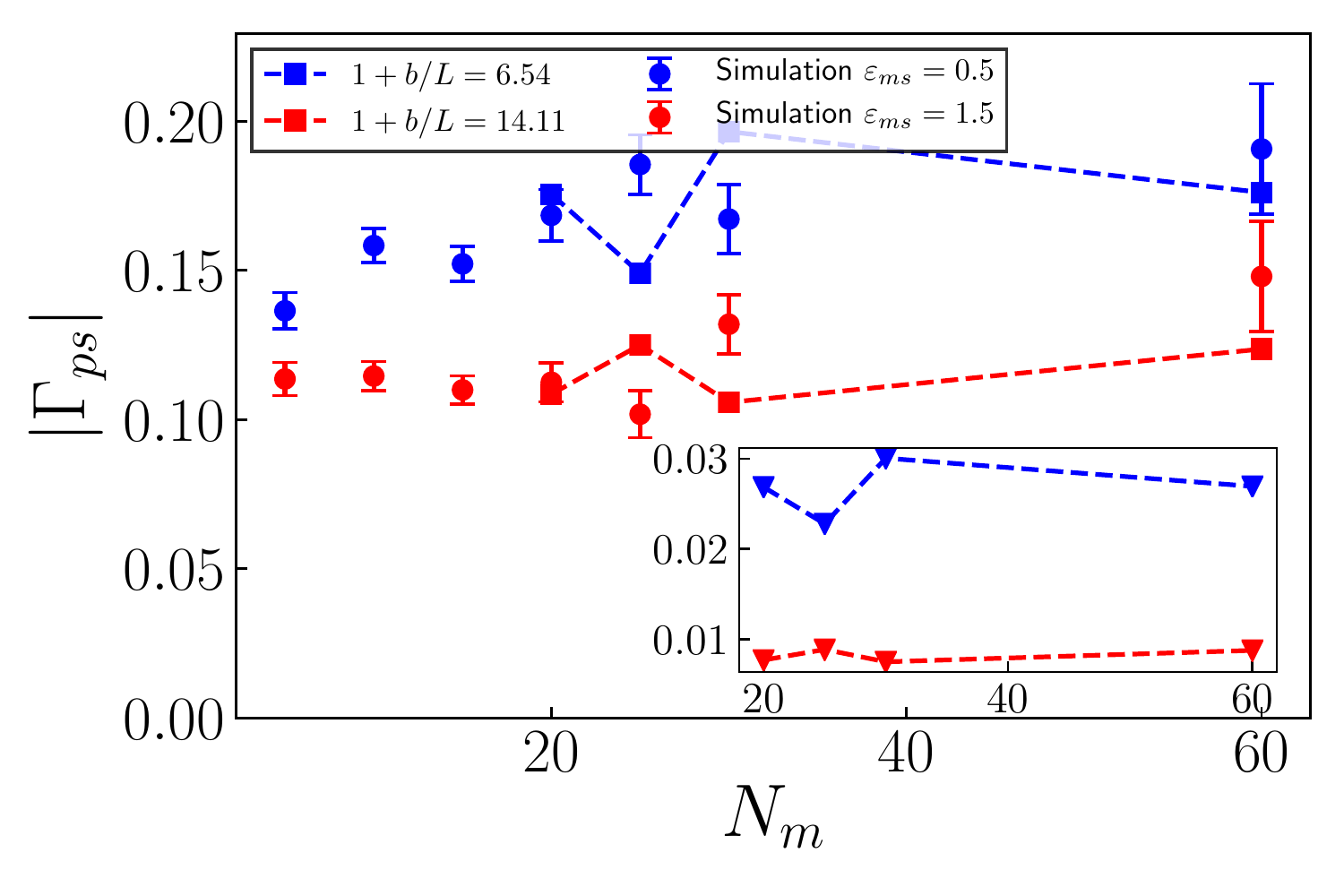}
\caption{Diffusiophoretic mobility $\Gamma_{ps}$ of a polymer  {\em vs} the number of monomers in the polymer $N_m$. The results $\epsilon^{LJ}_{ms} = 1.5$ are shown in red and in blue for $\epsilon^{LJ}_{ms} = 0.5$. The simulations results are represented as dots and the theoretical predictions (Eq.~\eqref{eq:dpvelocity}, including the amplification due to the hydrodynamic slip), are shown as squares. The insert shows the theoretical predictions without the slip correction. 
In the theoretical calculations  $a = R_h$, where $R_h$ is the hydrodynamic radius estimated, for the specific interaction $\epsilon_{ms}$ and thermodynamic conditions, using Stokes-Einstein relation (Eq.~\eqref{eq:hydroRadius}) }  
\label{fig:Mobility_N}
\end{figure} 

\begin{figure}[H]
\centering
\begin{subfigure}{0.49\linewidth}
  \caption{}
  \includegraphics[width=1\linewidth]{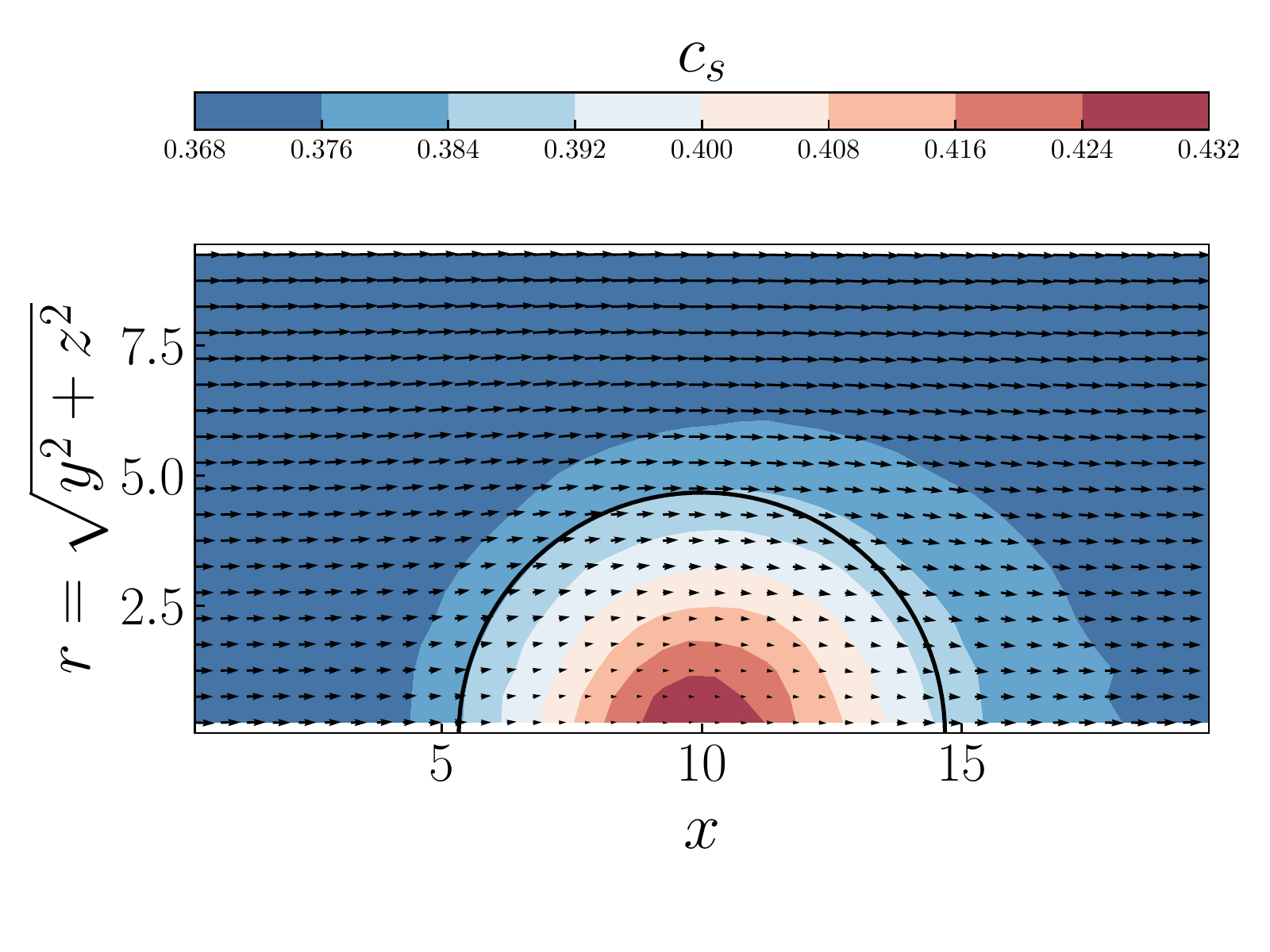}
  \label{fig:body_force}
\end{subfigure}%
\begin{subfigure}{0.49\textwidth}
    \caption{}
  \includegraphics[width=1\linewidth]{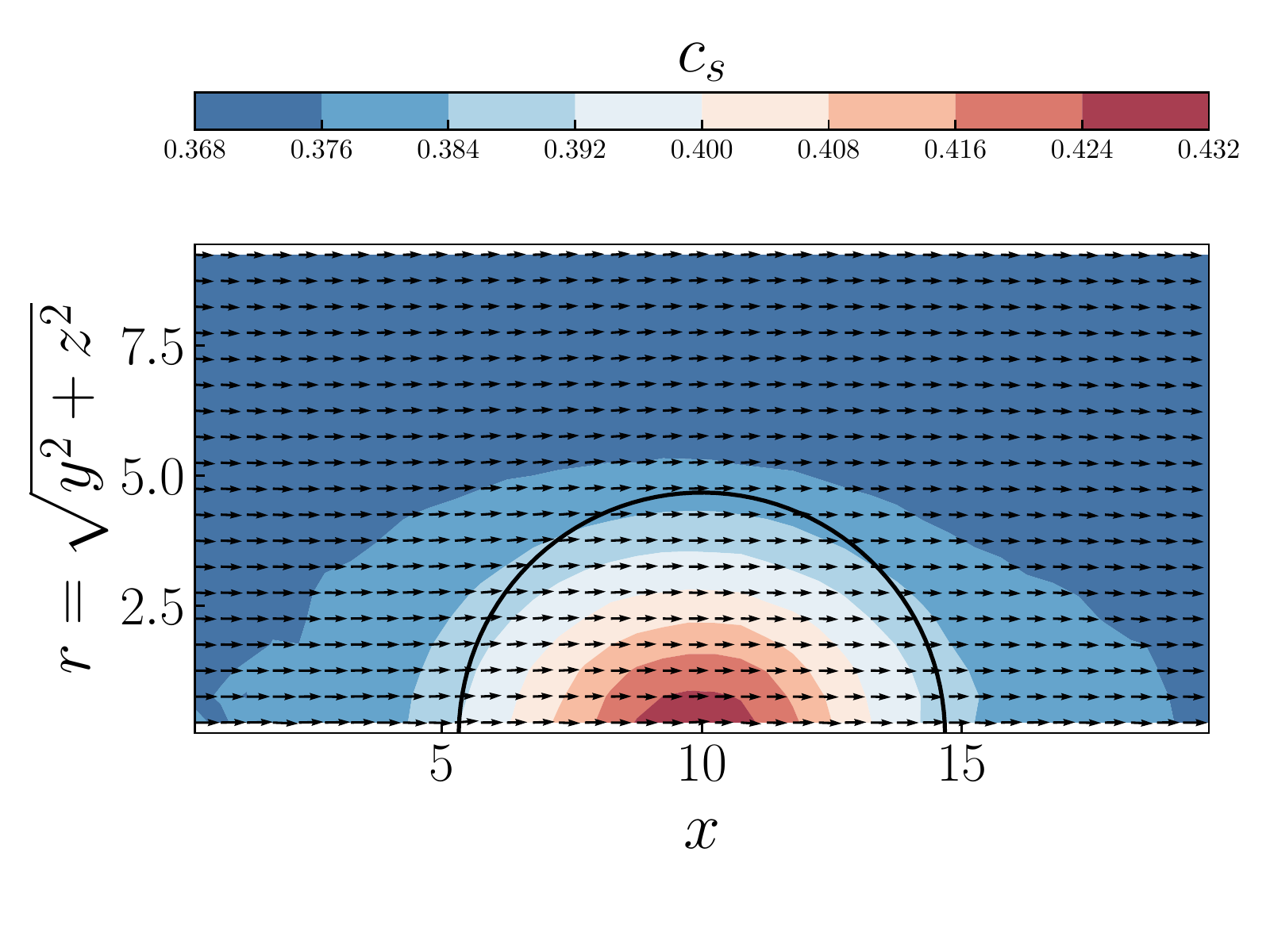}
  \label{fig:diffusio_phoretic}
\end{subfigure}
\caption{Flow around a polymer coil. (a) when a body force is applied and (b) for the diffusiophoretic case. The velocity field is measured in a coordinate system moving with the center of mass of the polymer. The black semicircle shows the equivalent colloid and the contours show the solute concentration, for both cases $\epsilon_{ms}=1.5$. The measurements were taken inside a cylinder with axis along the direction of the applied force passing through the center of mass of the polymer. The contours show the solute concentration $c_s$. }
\label{fig:flow_profiles}
\end{figure}

In Fig. \ref{fig:Mobility_N}, the simulation results are shown together with theoretical predictions replacing the polymer by an equivalent hard sphere with a radius $a=R_h$, with the hydrodynamic radius $R_h$ given by Stokes-Einstein relation (Comparison with the Kirkwood approximation for $R_h$ \cite{Kirkwood1954} are shown in SI ), 
\eq{\label{eq:hydroRadius}R_h\equiv\frac{k_b T}{6 \pi \eta D},}
where $\eta$ denotes the viscosity of the solution in the bulk, which was  computed independently, using the Green-Kubo expression relating $\eta$ to the stress auto-correlation function,   in an equilibrium simulation of the bulk fluid (see, e.g.~\cite{Hansen2006}). The diffusion coefficient $D$ was also computed from equilibrium simulations~\cite{Frenkel2002} taking into account the different interactions $\epsilon_{ms}$.
Fig. \ref{fig:Mobility_N} shows that  the diffusiophoretic mobility of the polymer increases with $N_m$. The large quantitative differences between the simulations and the theoretical approximations for a colloid with the same hydrodynamic radius are to be expected:  First of all, the assumption that the polymer coil behaves as a hard sphere with $a=R_h$ is rather drastic. To be more precise, this approximation (that was also used by Kirkwood \cite{Kirkwood1954}) assumes that the liquid molecules within the coil region move together, such that the whole assembly moves as a rigid sphere (see e.g.~\cite{Strobl2007}). This might be a good approximation for the diffusion of long polymer coils, but in the case of phoresis, it is unrealistic to assume that no solute/solvent can be transported through the polymer at distances less than $R_h$. The second (but related) questionable approximation is
 that $R_h$ defines the surface of the equivalent colloid in the integrals in Eqs.~\eqref{eq:zeroth}-\eqref{eq:second}. As a consequence, the contribution of any excess solute at a distance less than $R_h$ from the polymer center is ignored. As is clear from Fig.~\ref{fig:structuring} this assumption is incorrect and is likely to underestimate the real diffusiophoretic flow, in view of the fact that Fig.~\ref{fig:diffusio_phoretic} shows that there can be considerable solute advection for $r<R_h$. 
 
 Our simulations suggest that  better theoretical models for polymer diffusiophoresis are needed. In Fig.~\ref{fig:flow_profiles} we show the velocity field for a polymer with $N_m=30$ for two cases: when it is subjected to a body force i.e pressure gradient \ref{fig:body_force}  and under the influence of diffusiophoresis \ref{fig:diffusio_phoretic}. As is obvious from the figure, in both cases 
there is fluid motion within the polymer at distances less than  $R_h$ from its center of mass. However, there is an important difference between the flows inside the polymer for the pressure-driven and phoretically driven flows: strong hydrodynamic screening is found in the case of a pressure gradient while for diffusiophoresis, screening seems to be effectively absent. Notice that the density profile is somewhat asymmetric due to the  advection produced by the pressure-driven flow.

Shin {\em et al.} reported evidence for a similar absence of hydrodynamic screening in a dense plug of colloidal particles moving under the influence of diffusiophoresis~\cite{Shin2017b}. Shin {\em et al.} argued that the difference in screening in the case of phoretic flow, as opposed to flow due to body forces or pressure gradients, could be attributed to the difference in the range of the hydrodynamic flow fields in these two cases (($\sim1/r$) for body-forces and pressure driven flow, ($\sim1/r^3$) of phoretically induced flows~\cite{Anderson1989}. 
    
\section{Conclusions}

We have performed molecular dynamics simulation  on the diffusiophoresis of polymers in a fluid mixture under the influence of a concentration gradient of solutes.
In our non-equilibrium molecular dynamics simulation, we mimicked the effect of  an explicit concentration gradient in the system by imposing equivalent microscopic forces on the 
solute, solvent and monomers. This approach  allows us to use periodic boundary conditions and facilitates a systematic 
investigation of diffusiophoresis. 
Our results reveal a non-monotonic relation between the diffusiophoretic mobility
and the interaction strength between the polymer and the solute.
The findings imply  that, in the strong interaction regime, the phoretic mobility decreases with increasing monomer-solute interaction strength. 
This result can be understood by noting that  solutes that are strongly bound to the polymer  cannot contribute to diffusiophoresis.
Furthermore, we have demonstrated that the diffusiophoretic mobility of a (short) polymer cannot be explained in terms of a model that assumes that polymers behave like colloids with the same hydrodynamic radius.  Finally, we found effectively no screening of hydrodynamic flow inside a polymer moving due to diffusiophoresis, as opposed to what is observed in the case of a polymer that is moved through a fluid by an external force.

\section{Supplementary Material}
In the supplementary material we show the concentration distribution of solvent, solute, and solution as a function of radial coordinate from the center of mass of the polymer for $\epsilon_{ms} = 1.5$. We discuss simulations of a single, fixed colloid in a concentration gradient, and finally we show the diffusiophoretic mobilities of the polymer $\Gamma_{ps}$ vs the number of monomers in the polymer $N_m$ and compare with theoretical predictions based on the Kirkwood approximation for the  hydrodynamic radius.

\section{Acknowledgements}
This work was supported by the European Union Grant
No. 674979 [NANOTRANS]. DF and LB acknowledge support from the   Horizon 2020 program through 766972-FET-OPEN-NANOPHLOW. We would like to thank 
Richard P. Sear, Raman Ganti, Stephen Cox and Patrick Warren for the illuminating discussions.

\section{Data Availability}
The data that support the findings of this study are available from the corresponding author
upon reasonable request.

\bibliography{library}% Produces the bibliography via BibTeX.

\end{document}

\title[(Supplementary Material) Studying polymer diffusiophoresis with Non-Equilibrium Molecular Dynamics]{(Supplementary Material) Studying polymer diffusiophoresis with Non-Equilibrium Molecular Dynamics}
% Force line breaks with \\

\author{S.Ram\'irez-Hinestrosa}
\affiliation{Department of Chemistry, University of Cambridge, Lensfield Road, Cambridge CB2 1EW, United Kingdom}
\author{H. Yoshida}%
\affiliation{LPS, UMR CNRS 8550, École Normale Supérieure, 24 rue Lhomond, 75005 Paris, France}
\affiliation{Toyota Central R\&D Labs., Inc., Bunkyo-ku, Tokyo 112-0004, Japan}

\author{L. Bocquet}
\affiliation{LPS, UMR CNRS 8550, École Normale Supérieure, 24 rue Lhomond, 75005 Paris, France}

\author{D. Frenkel}
\email{df246@cam.ac.uk}
\affiliation{Department of Chemistry, University of Cambridge, Lensfield Road, Cambridge CB2 1EW, United Kingdom}

\date{\today}% It is always \today, today,

\maketitle
 \section{Concentration distributions}

\begin{figure}[H]
\centering
\includegraphics[scale=0.5]{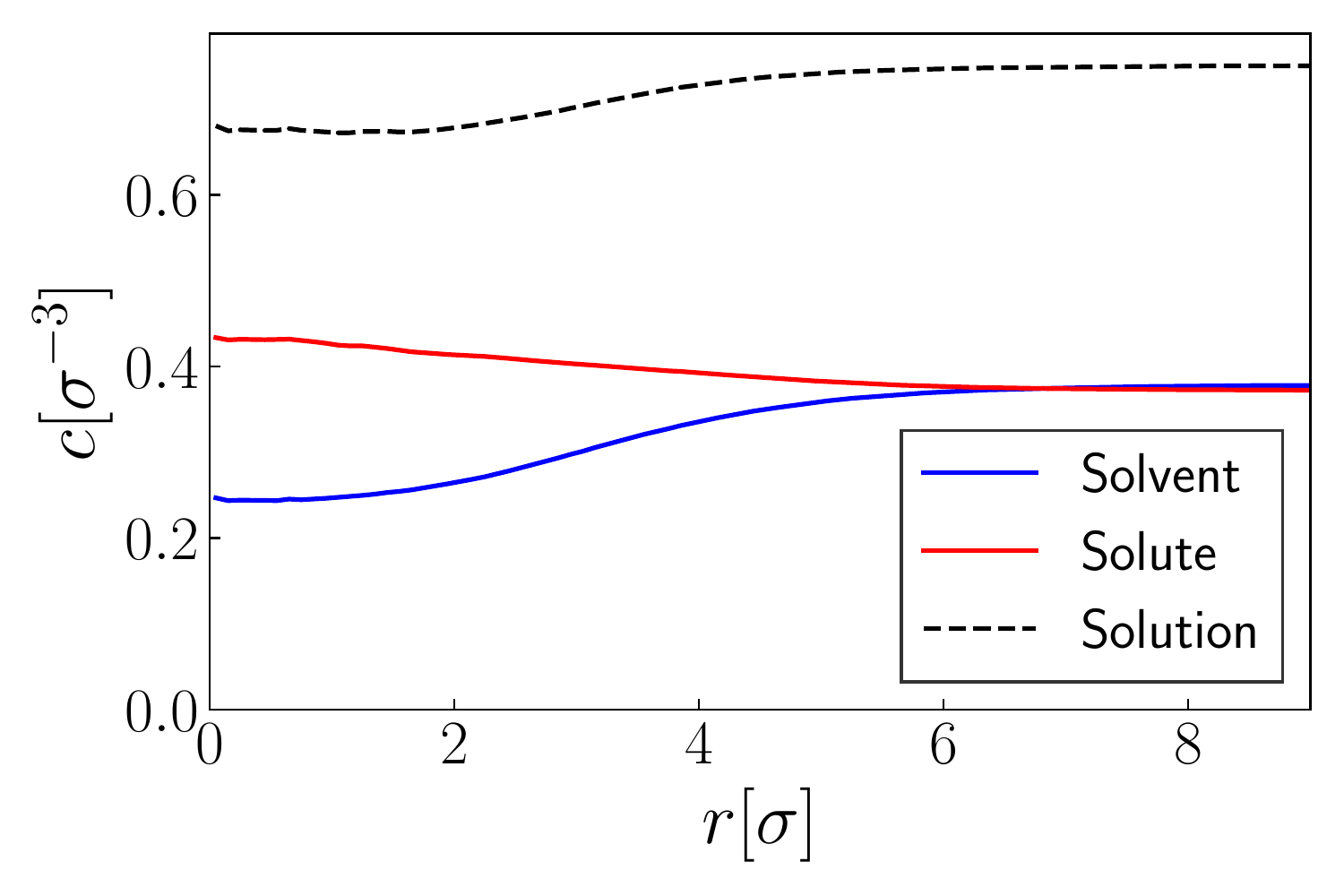}
\caption{Concentration of solvent, solute, and solution as a function of radial coordinate from the center of mass of the polymer for $\epsilon_{ms} = 1.5$.}  
\label{fig:con}
\end{figure}

In Fig. \ref{fig:con} we show the concentration distribution measured from the center of mass of the polymer in the radial direction for $\epsilon_{ms} = 1.5$. The figure shows that for $ r>7 \sigma_0$ the solution is unperturbed by the presence of the polymer. This also holds for all the $\epsilon_{ms}$ in the present study.

\section{Colloidal system simulations}

\begin{figure}[H]
\centering
\includegraphics[scale=0.7]{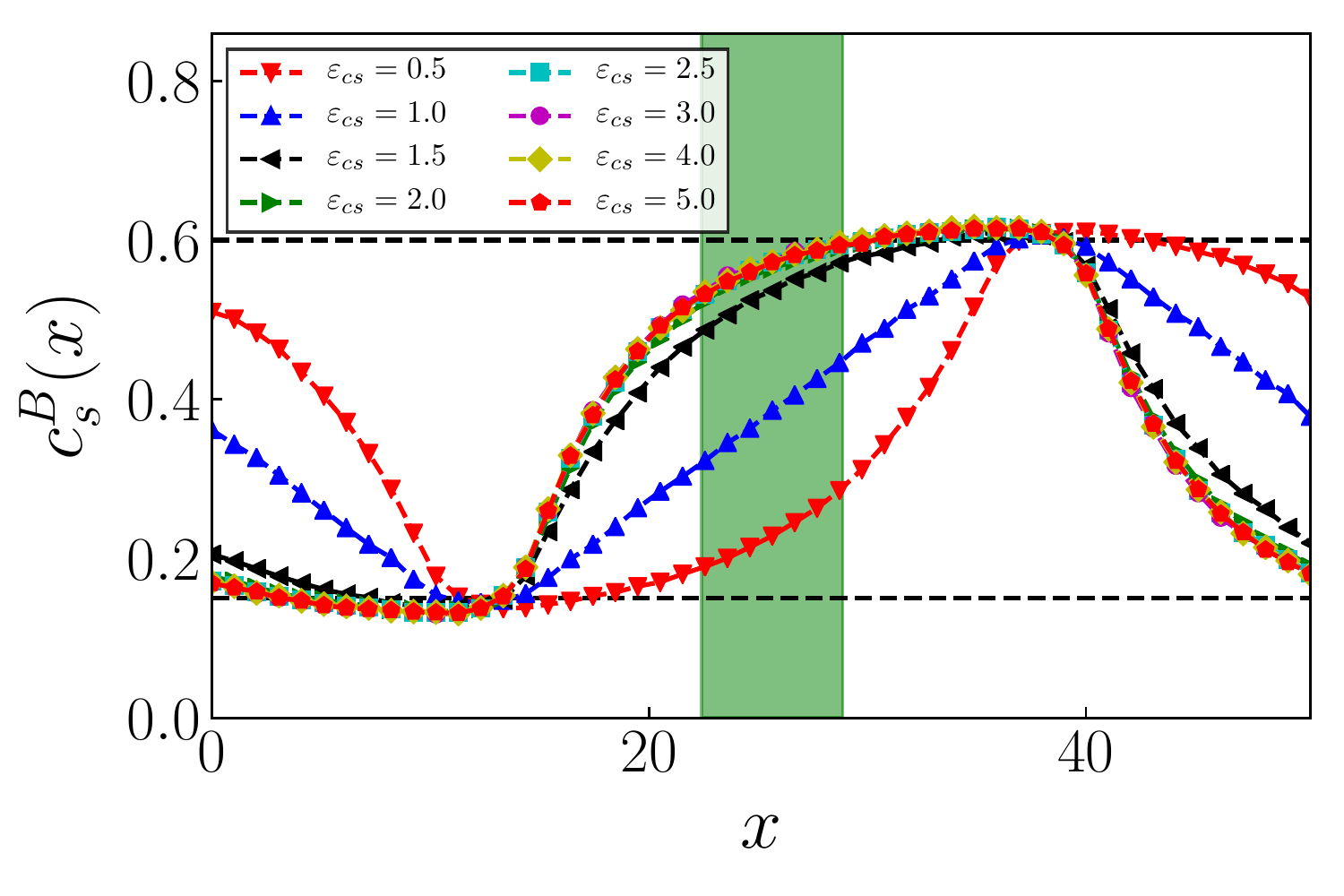}
\caption{Solute concentration profile in the case of a constant  imposed concentration difference. The figure shows the results for difference phoretic flow velocities, corresponding to different values of  $\varepsilon_{cs}$. 
We measure the concentration profiles at a lateral distance of at least 10$\sigma$ from the colloid, where the colloid does not directly perturb the concentration profile. 
The shaded region represents the $x$ position of the colloid and it is shown to emphasize the asymmetry in the concentration distribution created by advection.}  
\label{fig:conc_chunks}
\end{figure}

\begin{figure}[H]
\centering
\includegraphics[scale=0.5]{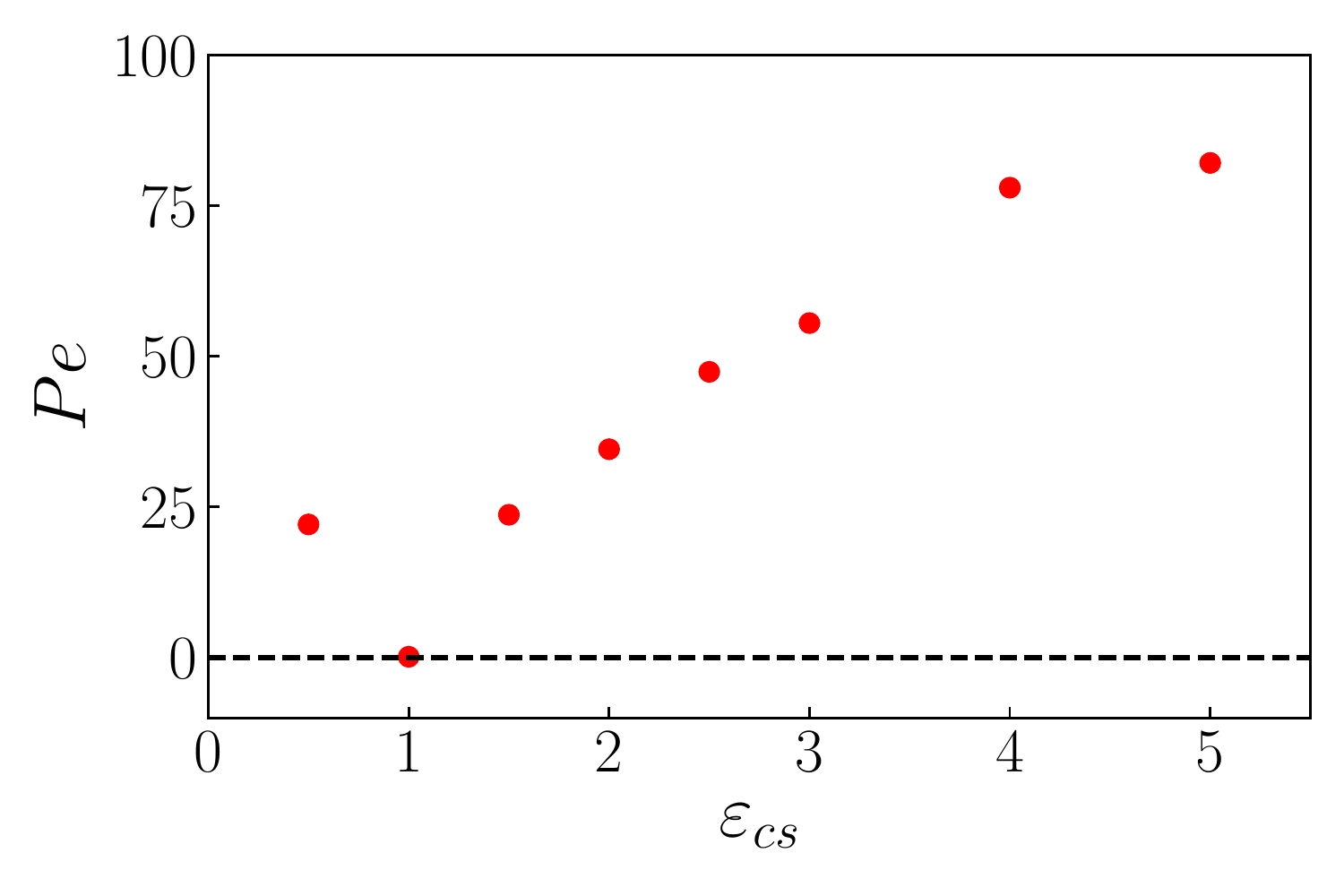}
\caption{P\'eclet number $Pe$ for the diffusiophoretic flow with several colloid-solute interaction strengths $\varepsilon_{cs}$ using explicit gradients. Note that, even for the smallest non-zero phoretic flow velocities $Pe>20$.}
\label{fig:Peclet}
\end{figure}

We performed simulations for a colloidal particle under the influence of both an explicit concentration gradient and the implicit method applying color forces. For the explicit system, the box size is (51.30 x 20.52 x 30.78) (in units of ${\sigma}_0$). A colloid was fixed in the center of the simulation box by placing a single Lennard-Jones particle with $\sigma_{cs} = \sigma_{cf}$ = 3.23 $\sigma_0 $, where the subscript $c$ denotes the colloid. The solution has the same properties as those described in the main text. The concentration gradient was imposed by using two-particle reservoirs,  with the concentration of solutes fixed at 0.6 $\sigma^{-3}$ in the high concentration reservoir and 0.15 $\sigma^{-3}$ in the low concentration one (see Fig. \ref{fig:conc_chunks}). The imposed concentration gradient is equivalent to $\nabla \mu_s \sim 0.06$ and is linear when there is no preferential interaction between the colloid and the solutes $\varepsilon_{cs} = 1.0$. As soon as phoresis starts, the concentration gradient becomes non-linear (in fact, exponential)  due to advection. As a result, the local concentration gradient at the location of the colloid is decreased (see also \cite{Weia}). The P\'eclet numbers for the solution in the bulk are shown in Fig \ref{fig:Peclet}

 \textbf{}

\begin{figure}[H]
\centering
\includegraphics[scale=0.3]{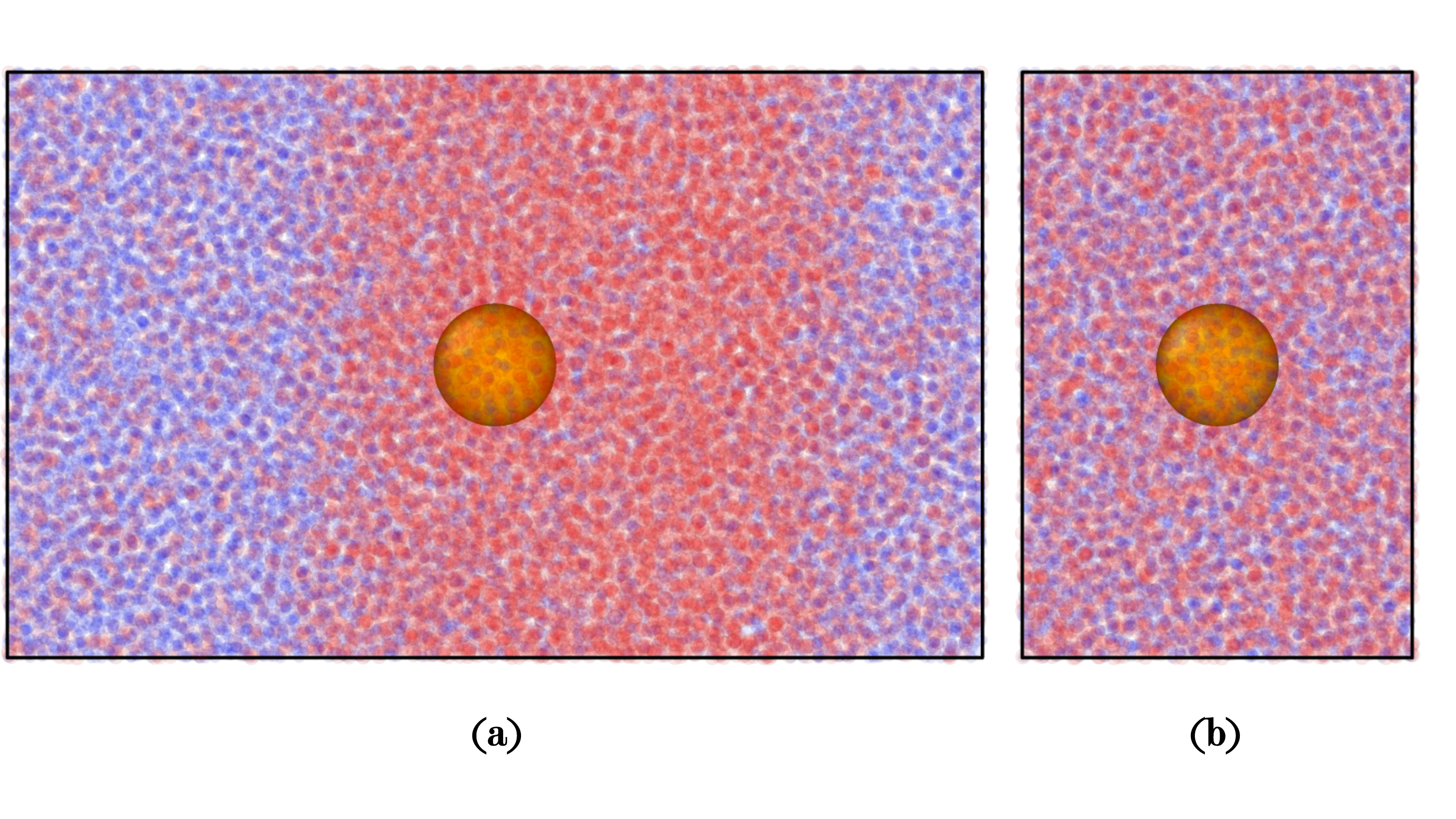}
\caption{Simulation boxes for (a) the explicit and (b) the implicit gradient systems  }  
\label{fig:snapshots}
\end{figure}

For the implicit method, we used a simulation box (20.52 x 20.52 x 30.78) (in units of ${\sigma}_0$)  with a colloid located in the center of the box. A snapshot of both systems is shown in Fig. \ref{fig:snapshots} Note that the deformation of the concentration profile in figure (b) (colour forces)  is barely visible. This observation illustrates the advantage of using colour forces rather than explicit gradients. In contrast, in Figure (a) (explicit gradient), the colloid is not in the center of the concentration profile (if it were, it would be at the red-blue boundary).

\begin{figure}[H]
\centering
\includegraphics[scale=0.6]{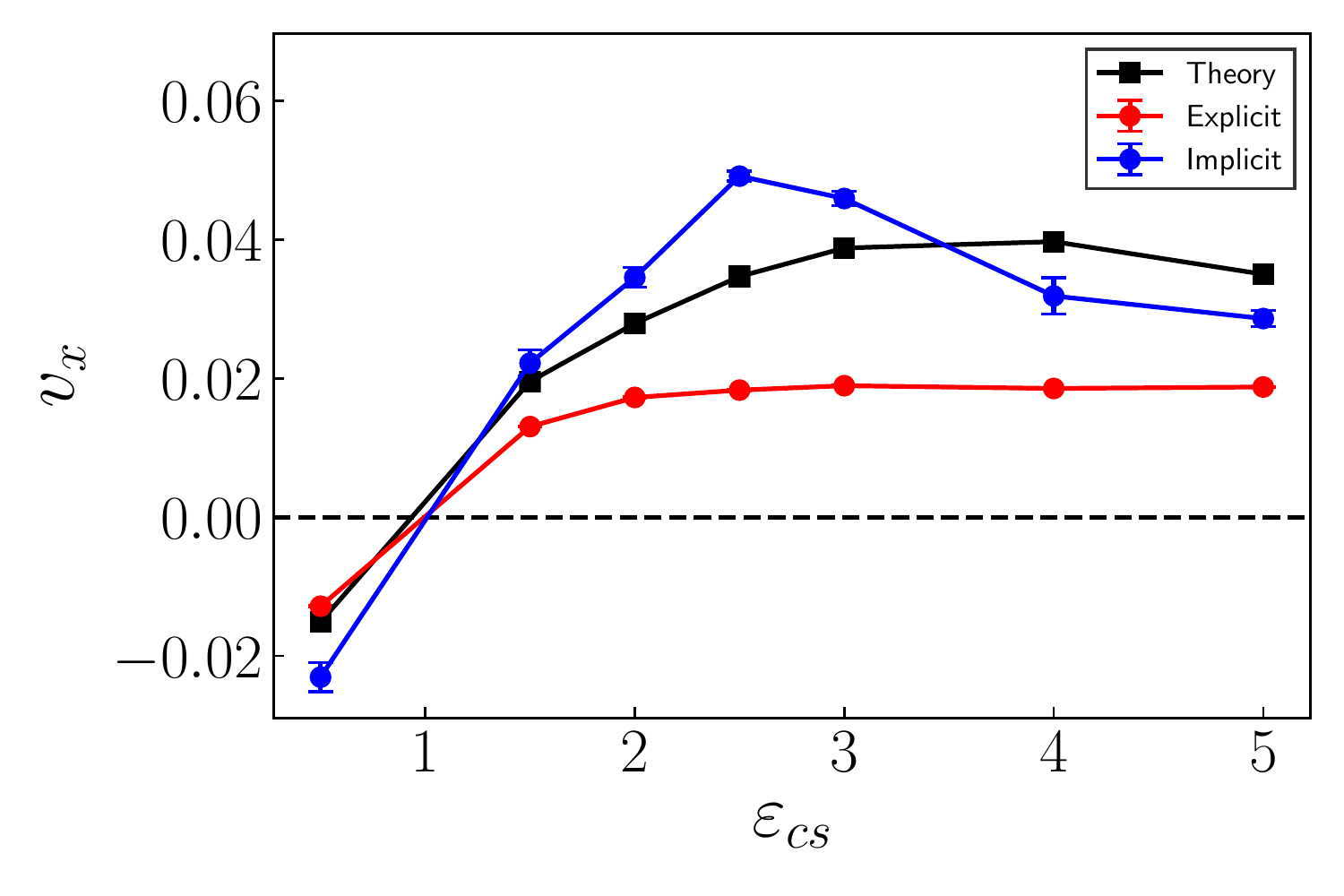}
\caption{Phoretic velocity $v_x$  with several colloid-solute interaction strengths $\varepsilon_{cs}$ obtained with explicit, implicit gradients and theoretical predictions.   }  
\label{fig:results}
\end{figure}

 In Fig.\ref{fig:results} we show the results obtained using the explicit and implicit gradients, together with the theoretical predictions using Anderson's theory including curvature corrections \cite{Anderson1982}. The results show that the explicit method is only suitable for small phoretic effects (small gradients or small $\varepsilon_{cs}$ as it is difficult to impose a linear gradient without using methods as described by Sharifi et al \cite{Sharifi-Mood2013} which artificially forces the concentration profile to be linear but at the expense of introducing (unphysical) sources/sinks of solute particles throughout the simulation box. 

To conclude, the advantages that the implicit gradient offers over the explicit gradient implementation are listed below:
\vspace{0cm}
\begin{myitemize}
    \item Allows simulating moving objects (polymers, colloids, etc) inside the concentration gradient. This would not be possible under periodic boundary conditions 
    \item  Only one parameter is required to change the chemical potential gradient, while in the explicit gradient method a new equilibrated system is required 
    \item There is no overhead due to exchange movements
    \item The system size is considerably smaller
    \item The explicit gradient method is only efficient for small P\'eclet number, otherwise the concentration gradient is disturbed by flow advection
\end{myitemize}
\newpage
\section{Diffusiophoretic mobilities using the Kirkwood approximation for $R_h$}

\begin{figure}[H]
\centering
\includegraphics[scale=0.7]{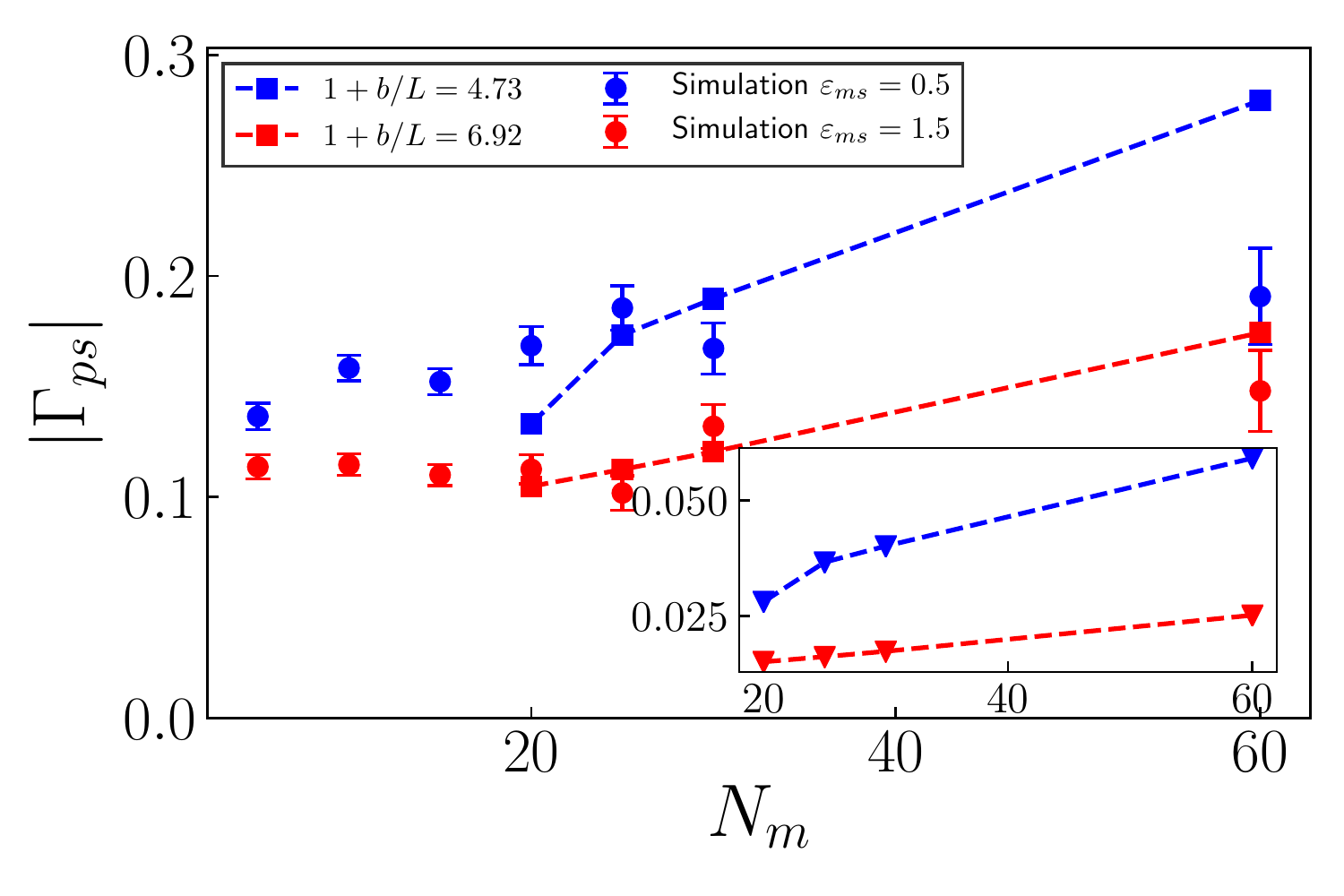}
\caption{Diffusiophoretic mobility $\Gamma_{ps}$ of a polymer  vs the number of monomers in the polymer $N_m$. The results are shown in red for $\epsilon^{LJ}_{ms} = 1.5$ and blue for $\epsilon^{LJ}_{ms} = 0.5$. The simulations results are presented as dots and the theoretical predictions using Anderson's prediction \cite{Anderson1982}, including the correction due to the hydrodynamic slip, are shown as squares. The insert shows the theoretical predictions without the correction for hydrodynamic slip. }  
\label{fig:kirkwood}
\end{figure}

In Fig. \ref{fig:kirkwood}, the simulation results are shown together with theoretical predictions replacing the polymer by an equivalent hard sphere with a radius $a=R_h$, with the hydrodynamic radius $R_h$ given Kirkwood's approximation of $R_h$ \cite{Kirkwood1954}. 

\bibliography{library}% Produces the bibliography via BibTeX.